\UseRawInputEncoding
\documentclass[runningheads]{llncs}
\usepackage[T1]{fontenc}
\usepackage{todonotes}
\usepackage{amsmath}
\usepackage{amssymb}
\usepackage{graphicx}

\usepackage{url}
\usepackage{tabularx}
\usepackage{appendix}
\usepackage{cite}
\usepackage{placeins}

\usepackage{color}          

\begin{document}
\title{Automated Reasoning for Vulnerability Management by Design}

\titlerunning{Automated Reasoning for Vulnerability Management by Design}
%
\author{Avi Shaked\inst{1} \orcidID{0000-0001-7976-1942} 
\and
Nan Messe\inst{2} \orcidID{0000-0002-3766-0710}
}
\authorrunning{Shaked et al.}
%
\institute{Department of Computer Science, University of Oxford, Oxford, OX1 3QD, UK\\
\email{avishakedse@gmail.com}
\and
IRIT, CNRS, UT2, France\\
\email{nan.messe@irit.fr}
}

\maketitle 
\begin{abstract}
For securing systems, it is essential to manage their vulnerability posture and design appropriate security controls. Vulnerability management allows to proactively address vulnerabilities by incorporating pertinent security controls into systems' designs. Current vulnerability management approaches do not support systematic reasoning about the vulnerability postures of systems' designs. To effectively manage vulnerabilities and design security controls, we propose a formally grounded automated reasoning mechanism. We integrate the mechanism into an open-source security design tool and demonstrate its application through an illustrative example driven by real-world challenges. 
The automated reasoning mechanism allows system designers to identify vulnerabilities that are applicable to a specific system design, explicitly specify vulnerability mitigation options, declare selected controls, and thus systematically manage vulnerability postures.
\end{abstract}
\section{Introduction}\label{sec:introduction}
Vulnerabilities in systems can be exploited by malicious actors, and are the primary, enabling factor of cyber security attacks~\cite{Matulevicius17}. Vulnerability management is therefore a critical aspect of secure system development and operation. Establishing the vulnerability posture of systems involves identifying and associating vulnerabilities with system components, and identifying the potential mitigation of these vulnerabilities by using security controls. Maintaining the vulnerability posture involves regularly updating the pertinent vulnerabilities and mitigation, to align with design decisions and the dynamic threat landscape (e.g., newly disclosed vulnerabilities affecting the system). Establishing and maintaining vulnerability postures allow the involved stakeholders (e.g., designers, risk managers, and
executives) to proactively identify, understand and mitigate potential
risks, consequently improving the
trustworthiness of systems~\cite{McGraw2006SoftwareSB,KHALIL2024103543,XIONG201953,ncsc_vm}.

In general, vulnerability management can relate to vulnerabilities found in the implementation of specific system constituents, which we relate to as \textit{implementation vulnerabilities}; as well as to conceptual classes of vulnerabilities (i.e., types of vulnerabilities), which we relate to as \textit{mechanism vulnerabilities}. Mitigating vulnerabilities at the mechanism level, rather than individually mitigating implementation vulnerabilities, is a desirable security-by-design approach to vulnerability management, because it addresses the root causes of potential security issues and prevents a wide range of existing as well as potential vulnerabilities and threats~\cite{cisa_sbd,almorsy2012supporting}.

The standard security
taxonomy provided by MITRE -- a global leader in systems engineering and cyber
security -- refers to mechanism vulnerabilities at various levels of
abstraction as \textit{weaknesses}, and to implementation vulnerabilities
simply as \textit{vulnerabilities}. The Common Weakness Enumeration (CWE)\footnote{\url{https://cwe.mitre.org/}, Accessed:
3/4/2025}
is a hierarchical organisation of weaknesses, while Common Vulnerabilities
and Exposures (CVE)\footnote{\url{https://www.cve.org/}, Accessed: 3/4/2025} lists implementation
vulnerabilities. Implementation vulnerabilities can manifest
weaknesses~\cite{huff2021towards,cisa_memsafe}. In this paper, we use
\textit{vulnerability} as a unified term to indicate either a mechanism issue or an implementation issue, unless otherwise specified.

Traditional vulnerability management approaches focus mainly on implementation vulnerabilities that are identified and addressed after system development. This reactive approach typically addresses individual implementation vulnerabilities but fails to address their underlying mechanisms -- mechanisms that may give rise to future implementation vulnerabilities. As a result, similar implementation vulnerabilities may remain unaddressed, leaving the system exposed to future attacks, sometime exploiting the same underlying mechanisms~\cite{1640171}. This requires to regularly check for new threats and design mitigation to address them~\cite{electronics12061333,khan2022systematic}.

Most approaches either focus narrowly on specific vulnerabilities or lack the scalability and rigor required to systematically integrate vulnerability management into system design processes~\cite{radanliev2024digital, 6502762, bdcc7010001}. In a comprehensive review of cyber security vulnerabilities and related concepts, Aslan et al. acknowledge the difficulty of ``\textit{applying domain knowledge for automated analysis}'' and list it as a major challenge alongside other related challenges, such as the time-consuming design of ``\textit{a secure system}'', the detection and prevention of unknown attacks, the protection of multiple components (as a system) and the increasing number of software vulnerabilities~\cite{electronics12061333}. Consequently, organisations struggle to adopt security-by-design principles, limiting their ability to ensure a robust, policy-driven security posture from the outset and proactively mitigate risks. A systematic literature review on security risks and practices calls for better ways of securing systems during development and, specifically, in its early stages~\cite{khan2022systematic}.

In recent years, security by design has received increasing attention~\cite{radanliev2024digital}. This is due to the premise that the earlier the security information is integrated into the system development lifecycle, the lower the overall debugging and maintenance costs incurred at later phases~\cite{meng2021verdict}. In previous work, we analyse current security-by-design approaches and their characteristics~\cite{BridgeSec_2024, messe2021security}. One of the challenges that remain is the lack of ability to rigorously reason about vulnerabilities at varying levels of abstraction during the design of a system. This challenge is also highlighted by other researchers~\cite{1640171, yskout2020}.

To address this challenge, there is a need for a formally specified reasoning mechanism that can support the association of vulnerabilities with specific system designs and reason about their mitigation, which includes the incorporation of relevant security controls into the designs. Such a mechanism should integrate seamlessly with existing design tools, enabling stakeholders to incorporate security considerations as part of the development process rather than as an afterthought.

In this paper, we present our work to develop a formally specified automated reasoning mechanism for vulnerability management by design and implement it as an extension of an open-source security modelling tool. We also illustrate how the mechanism is used to reason about the vulnerability posture of a system design. This illustration relies on real-world vulnerability data.

\section{Related Work}\label{sec:relatedwork}
Numerous studies have addressed vulnerability management. We overview some related work, highlighting how our approach complements existing methods and uniquely addresses the gap in scalable automated reasoning for vulnerability management. 

Some approaches suggest addressing the increasing number of reported vulnerabilities and the limited organisational capacity of remediation by prioritisation. A recent survey overviews dozens of approaches that attempt to prioritise vulnerabilities based on their exploitability~\cite{elder2024survey}. Like many of the survey approaches, Nowak et al. attempt to use the Common Vulnerability Scoring System (CVSS) to prioritise vulnerabilities for remediation, also taking into consideration the possible impact of their exploitation~\cite{nowak2023support}. Such attempts often involve the use of machine learning, which has limitations such as classification errors and need of representative data~\cite{electronics12061333}. Our work does not aim to prioritise vulnerabilities. Instead, we aim to identify vulnerabilities that are pertinent to a specific system design and reason about this in a way which allows to filter out the vulnerabilities that are already mitigated by existing security controls. 

Formal methods offer rigorous ways to reason about design characteristics, including security. 
Sengupta et al.~propose a formal methodology limited to detecting managerial vulnerabilities in enterprise information systems, without addressing technical system vulnerabilities~\cite{Sengupta2011}.
%
Fithen et al.~model vulnerabilities formally, using propositional
logic~\cite{fithen2004formal}. Their formalisation is limited to specific
product types (associated with \textit{Microsoft Windows}) and to
implementation vulnerabilities, with patching being the only
demonstrated security control and without addressing the temporal evolution
of the security landscape~\cite{TRADES_2023, electronics12061333}. Huff and Li propose a formal
approach to model software vulnerability risk in the context of the
network environment and firewall
configuration~\cite{huff2021towards}. Focused on operations, the approach
does not trivially translate into system design contexts, where system constituents are considered within their designated operating environment. To the best of our knowledge, there is no formally specified mechanism that allows reasoning to scale about the vulnerability posture of systems, while considering vulnerabilities at different levels of abstraction and their potential mitigation.


Other approaches lack formal foundations. For example, the approach proposed by
Longueira-Romero et al.~assesses known vulnerabilities in industrial components using directed graphs, without any formalisation of their mitigation~\cite{Longueira_Romero_2022}. Almorsy et al. use the declarative language OCL to automate limited aspects of vulnerability
analysis, without providing formal definitions~\cite{almorsy2012supporting}. 

Rouland et al. propose a model-driven formal approach to specifying mechanism vulnerabilities and controls with respect to software architecture~\cite{rouland2025model}. The implemented vulnerabilities library is limited, and the approach does not take into consideration implementation vulnerabilities and, accordingly, does not integrate with common vulnerability management practices (e.g., addressing CVEs via patching). This can be seen as complementary to our work, as it allows to formally define mechanism vulnerabilities, while our more conceptual approach aims to curate such information within the context of existing security design practices and body of knowledge and with respect to high-level system design.

TRADES Tool~\cite{TRADES_2023} is an open-source systems security design tool. It is underpinned by a semi-formal modelling methodology and relies heavily on a domain metamodel. We have previously extended TRADES Tool with vulnerability management concepts~\cite{BridgeSec_2024}. However, the lack of formal foundations for addressing vulnerability management by design -- including the hierarchical organisation and mitigation of vulnerabilities -- remains a significant gap. In this paper, we present our work to develop a formally specified automated reasoning mechanism for vulnerability management by design and implement it as an extension of TRADES Tool. 


\section{Formally Specified Automated Reasoning for Vulnerability Management}\label{sec:vul_man_for}
We provide a formally specified, automated reasoning mechanism to analyse the vulnerability
posture of systems designs. The contribution of this mechanism is two-fold. First, it associates potential vulnerabilities with system components based on pre-specified associations between vulnerabilities and component types. Second, it assesses the vulnerability posture of the design by determining whether
the security controls associated with the components are sufficient to
mitigate the associated vulnerabilities. The latter is performed according
to pre-specified rules that provide guidance for the mitigation of
vulnerabilities -- in the context of specific component types -- by
suggesting pertinent controls.

We employ the following finite sets to formalise concepts related to
vulnerability management:
\[\begin{array}{@{}l@{\quad\textrm{--}\quad}l@{}}
C & \textrm{the components $c \in C$ available in a system design or threat model} \\
T & \textrm{the component types $t \in T$ available} \\
V & \textrm{the known vulnerabilities $v \in V$ available} \\
S & \textrm{the security controls $s \in S$ available} \\
R & \textrm{the security rules $r \in R$ available} 
\end{array}\]


\noindent To obtain a formal vulnerability management model, we would define the
following mappings: $\mathrm{VULNS} : T \to \mathcal{P}(V)$ maps a
component type to the vulnerabilities that affect it; $\mathrm{TYPES} : C
\to \mathcal{P}(T)$, maps a component to the component types it manifests;
$\mathrm{CONTROLS} : C \to \mathcal{P}(S)$, maps a component to the
controls associated with it; $\mathrm{AVULNS} : V \to \mathcal{P}(V)$,
maps a vulnerability to vulnerabilities of higher abstraction that it
manifests, in a unidirectional, acyclic manner; $\mathrm{RVULNS} : R \to \mathcal{P}(V)$, maps a rule to the
vulnerabilities to which it applies; $\mathrm{RTYPES} :R \to
\mathcal{P}(T)$, maps a rule to the component types to which it applies;
$\mathrm{RCONTROLS} : R \to \mathcal{P}(S)$, maps a rule to the security
controls it suggests as mitigation.

In our framework, these sets and mappings can be manually defined and/or automatically generated, without loss of generality or limitation. The result is a body of formally codified knowledge about the system design and security at a particular point in time, which can then be used as the basis
for reasoning about the vulnerability posture of the system. Evolution of the design -- for example adding components or augmenting the mitigations in place -- can be done by augmenting the relevant mappings. Likewise, new
vulnerabilities or rules can be added by augmenting the relevant sets and mappings. The reasoning can then be performed by an automated reasoning mechanism, as further explained. 

As part of our vulnerability management reasoning mechanism, we wish to derive an indication whether a specific component is vulnerable. First, the reasoning mechanism can collect all the vulnerabilities that apply
to a component $c$ $\in$ $C$ based on the component's types:
\begin{equation*}
\mathrm{CVULNS}(c) = \bigcup_{\mkern-37mu t \in \mathrm{TYPES}(c)\mkern-37mu} \mathrm{VULNS}(t)
\end{equation*}

\noindent The mechanism can then check whether each vulnerability so obtained is
mitigated by a collection of security controls that is deemed appropriate by some security rule.
%
For a given vulnerability $v$ of a component $c$, 
a rule is pertinent if one of its types is also a type possessed by 
$c$ and applies to $v$. Proper mitigation for the
vulnerability is considered as a situation in which all the security controls $s$ identified by the rule are associated with the component. Formally:
\[
\mathrm{MitigatedV}(v,c) \equiv  \exists r \in R.\, 
\left \{\begin{array}{@{}l}
\exists t \in \mathrm{RTYPES}(r).\, t\in \mathrm{TYPES}(c) \; \wedge \\[1mm]
v \in \mathrm{RVULNS}(r) \; \wedge \\[1mm]
\forall s \in \mathrm{RCONTROLS}(r).\, s \in \mathrm{CONTROLS(c)}
\end{array}\right.
\]
$\mathrm{MitigatedV}$ defines whether a vulnerability is directly mitigated 
by a relevant control. But a vulnerability can also be mitigated indirectly, by
mitigating all the higher level abstractions it manifests. Formally, this is expressed by
\[
\mathrm{Mitigated}(v,c) \equiv 
\left \{\begin{array}{@{}l}
  \mathrm{MitigatedV}(v,c) \; \vee \\[1mm]
   (\mathrm{AVULNS}(v) \neq \varnothing \wedge \forall v' \in \mathrm{AVULNS}(v).\, \mathrm{Mitigated}(v',c))
  \end{array}\right.
\]
\noindent where the well-foundedness of this recursive formulation is
established by the fact that the abstraction relation on vulnerabilities is a finite partial order.

We can now define a predicate that allows the reasoning mechanism to indicate whether a component has a vulnerability that remains unaddressed by the associated controls:

\[
\mathrm{Vulnerable}(c) \equiv\ \exists v \in \mathrm{CVULNS}(c).\, \neg \text{Mitigated}(v,c)
\]
\noindent Finally, the automated reasoning mechanism can make sure that the design model satisfies the following property, indicating that no unmitigated vulnerabilities exist in any of the components in the design model:
\begin{property} \label{Property1}
$\forall c \in C.\, \neg \mathrm{Vulnerable}(c)$.

\end{property}

    
\section{Implementation}\label{sec:trades_int}

\begin{figure}[h]
\centering
\includegraphics[scale=0.12]{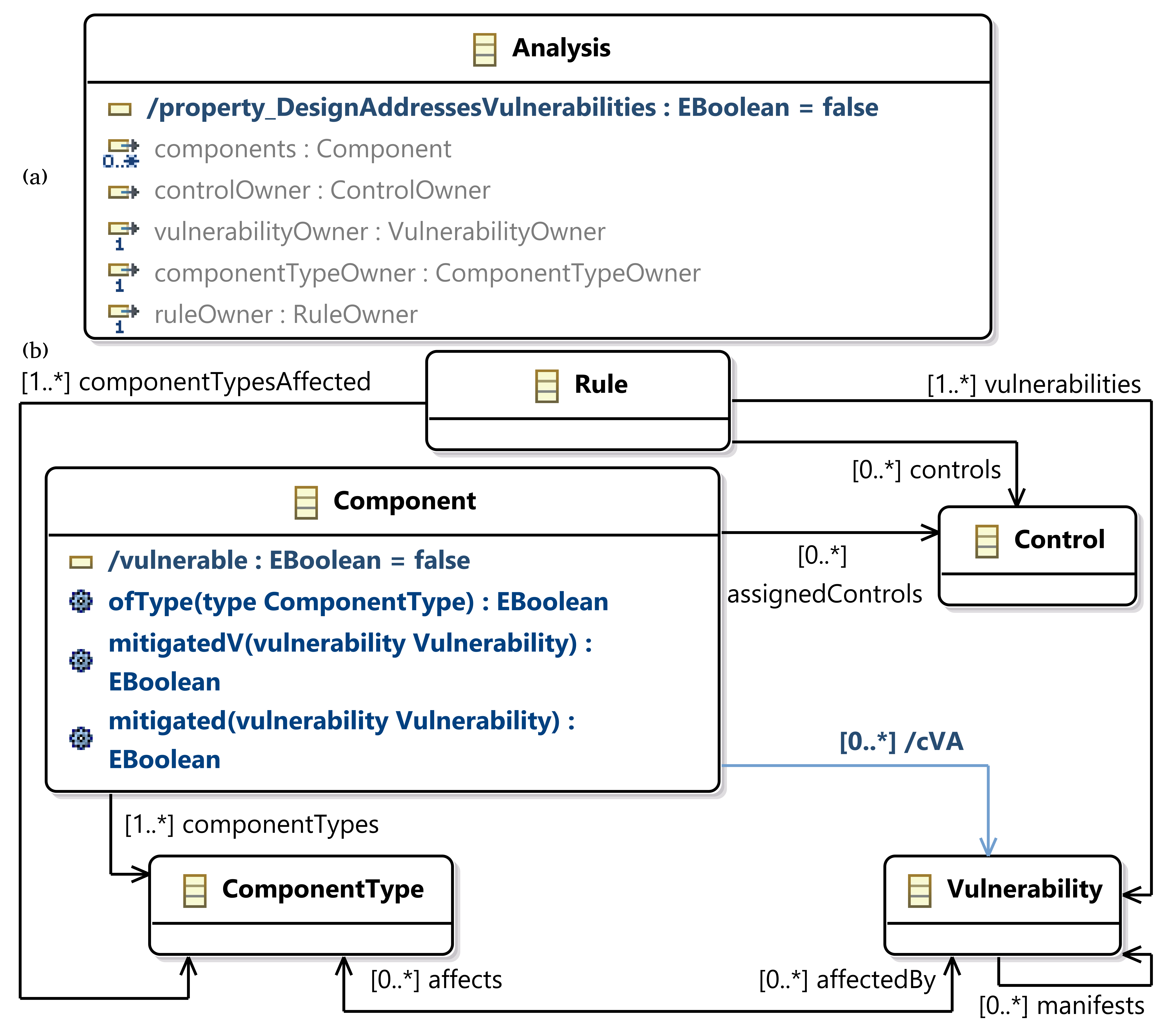}
\caption{TRADES Tool metamodel excerpt, showcasing integration of the formally-specified vulnerability analysis mechanism.}\label{fig2}
\end{figure}

Fig.~\ref{fig2} shows an excerpt from the \textit{metamodel} of the
TRADES Tool, into which the above vulnerability management concepts and automated reasoning mechanism are integrated. The metamodel has five classes that correspond to the five sets introduced above: $C$, $T$, $S$,
$V$ and $R$ (Fig~\ref{fig2} (b)). For example, {\bf\textsf Rule} corresponds to the set $R$ and
{\bf\textsf Control} corresponds to the set $S$. The labeled arrows
represent the family of mappings introduced above, with the label giving
the name of the mapping in TRADES Tool. For example, {\bf\textsf componentTypes} from {\bf\textsf Component} to {\bf\textsf ComponentType}
represents the mapping $\mathrm{TYPES} : C \to \mathcal{P}(T)$.

In TRADES Tool, the {\bf\textsf Analysis} concept ((Fig~\ref{fig2} (a)), is the root
element of a design model and can store instances of the other concepts/classes.
The reasoning mechanism implementation adds derived elements, annotated in
blue and with a slash prefix. These include: the {{\bf\textsf cVA} relation
between {\bf\textsf Component} and {\bf\textsf Vulnerability},
corresponding to the formalism's $\mathrm{CVULNS}$ mapping; and the
derived attributes {\bf\textsf vulnerable} for the {\bf\textsf Component}
and \newline {\bf\textsf {property\_DesignAddressesVulnerabilities}} for the
{\bf\textsf Analysis}, corresponding to the $\mathrm{Vulnerable}$
predicate and \textit{Property~\ref{Property1}}, respectively. Operations
are also added to the {\bf\textsf Component}, implementing the
$MitigatedV$ and $Mitigated$ predicates.

\section{Illustrative example}\label{sec:example}
We describe an illustrative example application using the formally-specified automated reasoning mechanism for vulnerability management and its TRADES Tool implementation. The TRADES Tool extension is available from the TRADES Tool Github repository\footnote{\url{https://github.com/UKRI-DSbD/TRADES}}. The example application TRADES
model is also available online\footnote{\url{https://github.com/UKRI-DSbD/TRADES/tree/VulManEx/VulManEx}}.

Consider the following case of assessing the vulnerability posture of a computer system being developed for the internal use of an
organisation. According to the preliminary design, the system comprises two software components: a UNIX-like operating system instance, and an organisational application instance. The set of types for the system's components accordingly includes two types: ``UNIX-like operating system'' and ``internally developed application''.

For the preliminary design, we -- as security engineers -- identify the
vulnerability \textit{Improper Restriction of Operations within the Bounds of a Memory Buffer (CWE{-}119)} as being of interest. A recommended
mitigation for such vulnerabilities is to \textit{use memory-safe programming languages}~\cite{cisa_memsafe}\footnote{Also: \url{https://cwe.mitre.org/data/definitions/119.html}, Accessed:
3/4/2025}. Accordingly, we
design a security rule, called ``rule1'', to address this vulnerability in
internally-developed applications, mandating the recommended security
control. The formalisation of this design situation defines the following sets:
\[
\begin{array}[t]{@{}l@{\;=\;}l@{}}
C & \{\mathrm{OperatingSystem}, \mathrm{Application}\} \\
T & \{\mathrm{UNIX\_like\_operating\_system}, \mathrm{internally\_developed\_application}\} \\
V & \{\mathrm{CWE\textrm{-}119}\} \\
S & \{\mathrm{use\_memory\_safe\_languages\}} \\
R & \{\mathrm{rule1}\}
\end{array}
\]

\noindent and these mappings:
\[
\begin{array}[t]{@{}l@{\;=\;}l@{}}
  \mathrm{VULNS}(\mathrm{internally\_developed\_application})
   & \{\mathrm{CWE\textrm{-}119}\}\\
  \mathrm{TYPES}(\mathrm{OperatingSystem})
  & \{\mathrm{UNIX\_like\_operating\_system}\}\\
  \mathrm{TYPES}(\mathrm{Application})
  & \{\mathrm{internally\_developed\_application}\}\\
  \mathrm{RVULNS}(\mathrm{rule1})
  &\{\mathrm{CWE\textrm{-}119}\} \\
  \mathrm{RTYPES}(\mathrm{rule1})
  & \{\mathrm{internally\_developed\_application}\} \\
  \mathrm{RCONTROLS(\mathrm{rule1})}
  & \{\mathrm{use\_memory\_safe\_languages}\}
\end{array}
\]
Fig.~\ref{figA1} shows the various model elements and highlights
$\mathrm{rule1}$ in TRADES Tool. We note that TRADES Tool extension that we developed allows to import the entire CWE catalogue of mechanism vulnerabilities and their hierarchical organisation into our design workspace. Once this is done, the catalogue can be used as a resource for vulnerabilities.

\begin{figure}[h]
\centering
\framebox{\includegraphics[scale=0.45]{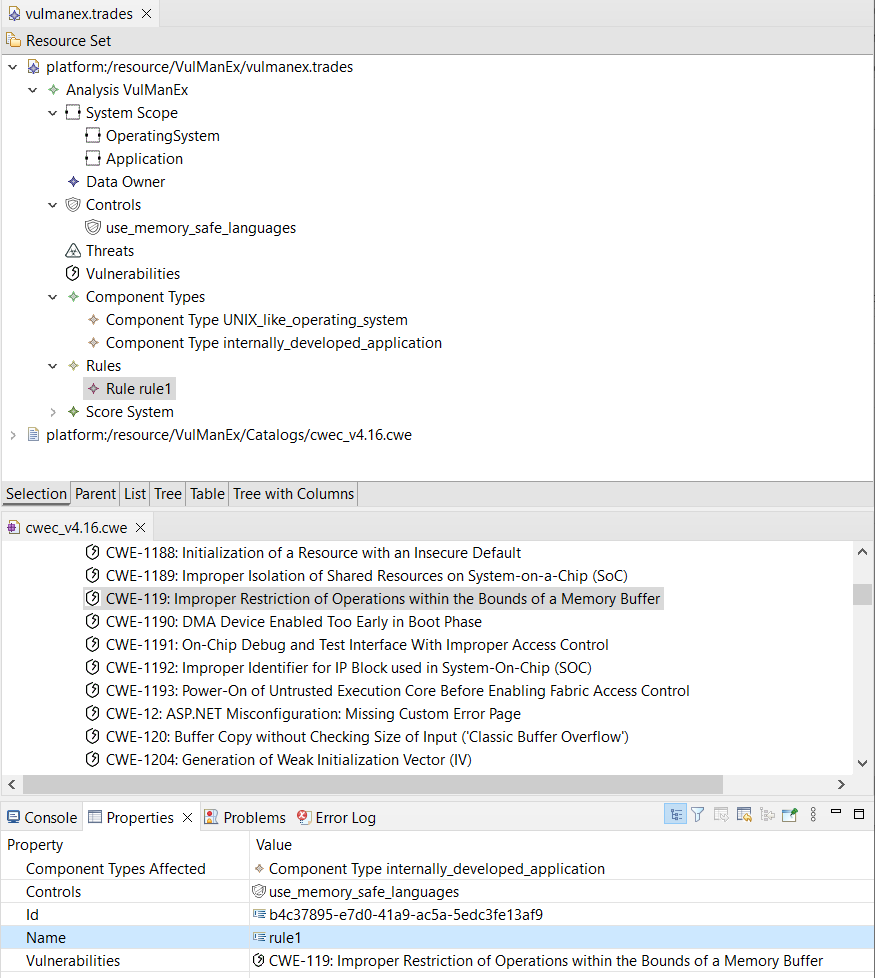}}
\caption{The illustrative example in TRADES Tool: the Resource Set panel shows the model elements grouped by their concepts, the middle panel shows an excerpt from the imported CWE catalogue, and the Properties panel shows the attributes and mappings of $\mathrm{rule1}$.} \label{figA1}
\end{figure}

At this point in the development life cycle, no controls have been assigned to any component. If we exercise the automated reasoning mechanism, \textit{Property~\ref{Property1}} is violated, with $\mathrm{Vulnerable}\mathrm{(Application})$ being true, with a counterexample being $\neg \mathrm{Mitigated}(\mathrm{CWE\textrm{-}119}, \mathrm{Application})$. Fig.~\ref{figA2} shows the initial $\mathrm{Vulnerable}$ status of the $\mathrm{Application}$ component (at the very bottom), alongside other mappings, including the unmitigated vulnerabilities mapping to CWE-119, which is automatically set by the automated reasoning mechanism as a counterexample.

To deal with the vulnerable component, we can use the knowledge codified in $\mathrm{rule1}$ to assign a security control
as a requirement to the Application component, setting
$\mathrm{CONTROLS}(\mathrm{Application})$ to be 
$\{\mathrm{use\_memory\_safe\_languages}\}$.  If we now re-apply the reasoning mechanism, \textit{Property~\ref{Property1}} is satisfied, i.e., all of the specified components are not vulnerable.  Fig.~\ref{figA3} shows the assignment of the
$\{\mathrm{use\_memory\_safe\_languages}\}$ control to the $\mathrm{Application}$ component, resulting in the component being assessed as not vulnerable (based on $\mathrm{rule1}$).
\begin{figure}[hpt]
\centering
\framebox{\includegraphics[scale=0.55]{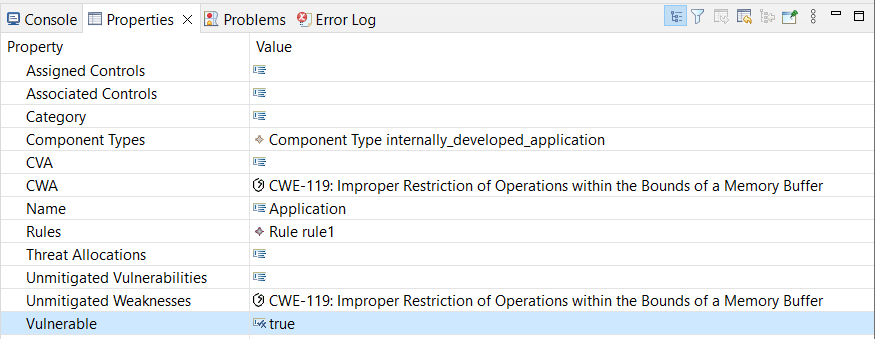}}
\caption{The Application component assessed as Vulnerable.} \label{figA2}
\end{figure}

\begin{figure}[!hpt]
\centering
\framebox{\includegraphics[scale=0.6]{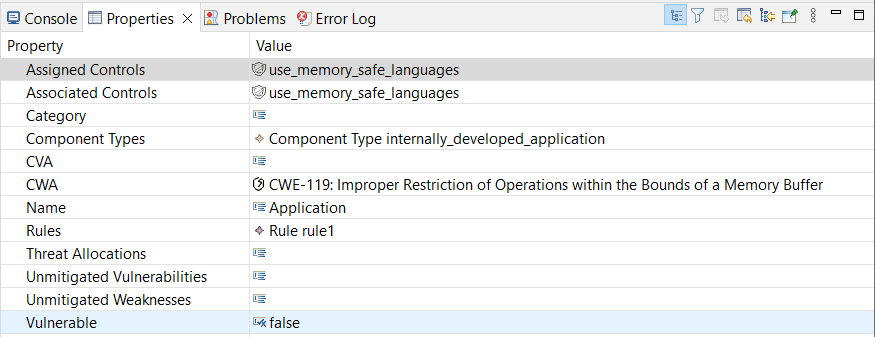}}
\caption{The Application component assessed as not Vulnerable due to the assigned control.} \label{figA3}
\end{figure}

As our system design evolves, we identify the specific operating system we wish to use -- FreeBSD version 14. Common Platform Enumeration (CPE)\footnote{\url{https://cpe.mitre.org/}} is a structured naming scheme for information technology systems, software, and packages. CPE is useful in associating vulnerabilities with affected software. Accordingly, we introduce a new component type, with the CPE of FreeBSD version 14: \textit{cpe:2.3:o:freebsd:freebsd:14.0:-:*:*:*:*:*:*}. This is shown in Fig.~\ref{figBSDtype}. We can now query the NIST's National Vulnerability Database (NVD) to retrieve implementation vulnerabilities associated with the newly incorporate component type. For conciseness and readability, and to reduce cognitive load on the reader, we query the NVD only for vulnerabilities that manifest the aforementioned CWE-119, which results in two CVE records: CVE-2011-2895 and CVE-2020-10565 (Fig.~\ref{figNVDq1}). These vulnerabilities are then imported into our design space, along with their associations -- as detailed in the NVD database -- with mechanism vulnerabilities and with the queried component type (Fig.~\ref{figVul}).


\begin{figure}[!hpt]
    \centering
    \hspace{-4cm}
    \begin{minipage}[t]{0.55\textwidth}
        \centering
        \includegraphics[scale=0.4]{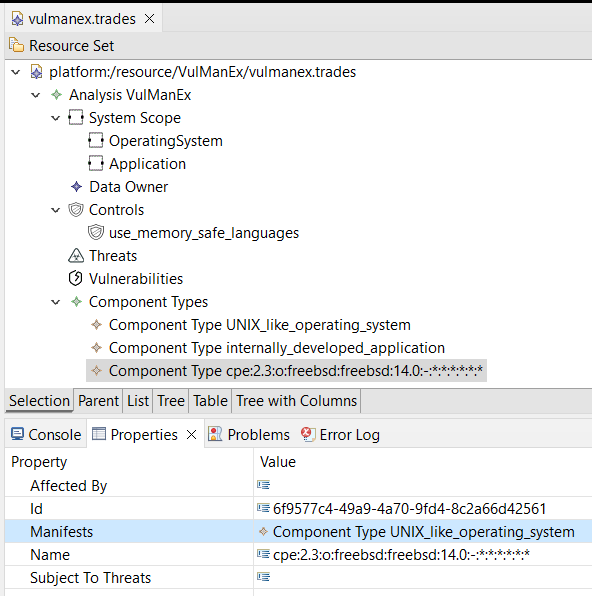}
        \caption{A new CPE-based component type is incorporated into our design space.}
        \label{figBSDtype}
    \end{minipage}
    \hspace{2cm}    
    \begin{minipage}[t]{0.48\textwidth} 
        \centering
        \includegraphics[scale=0.3]{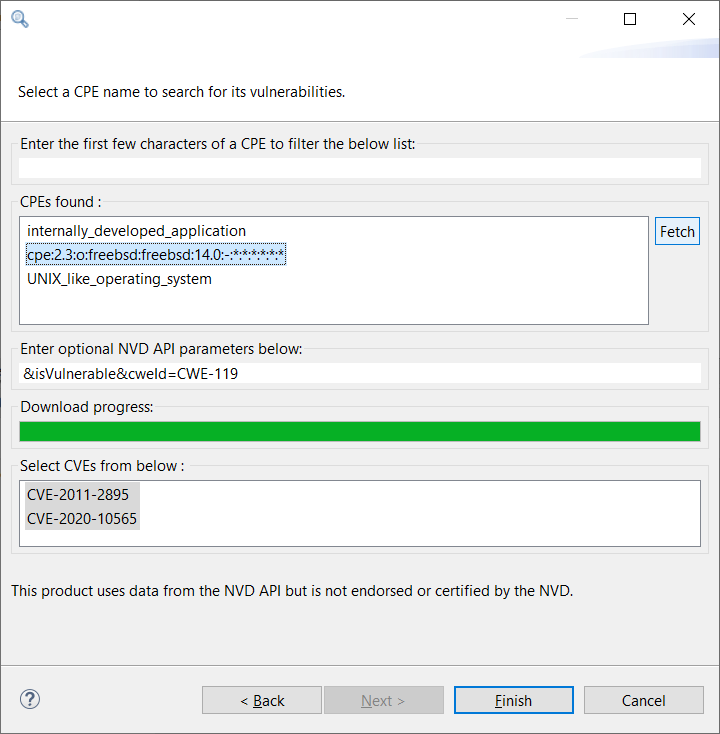}
        \caption{Querying NVD for CWE-119 related vulnerabilities in FreeBSD 14.}
        \label{figNVDq1}
    \end{minipage}    
\end{figure}


\begin{figure}[!hpt]
\centering
\hspace*{-4cm}
\framebox{\includegraphics[scale=0.4]{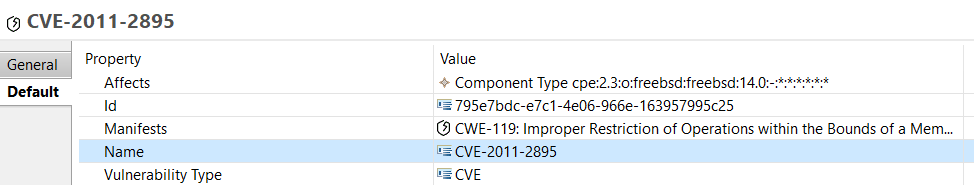}
{\includegraphics[scale=0.35]{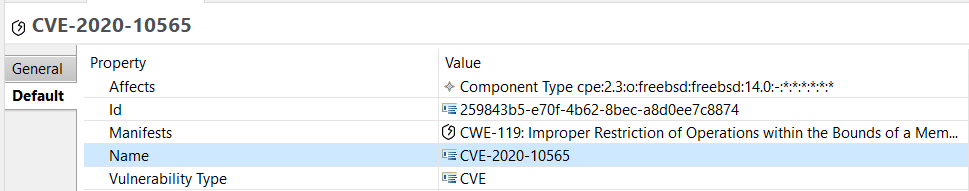}}}
\caption{Imported implementation vulnerabilities (CVEs) and their association with mechanism vulnerabilities (CWEs) and component types.} \label{figVul}
\end{figure}

Fig.~\ref{figOStype} shows the consequences of setting the type mapping of the OperatingSystem component to include the new FreeBSD 14.0 component type. Automatically, the reasoning mechanism deduces that the component is vulnerable, and lists the unmitigated vulnerabilities (the two CVEs that were imported into our design space).
\noindent
\begin{figure}[!hpt]
\centering
\framebox{\includegraphics[width=\textwidth]{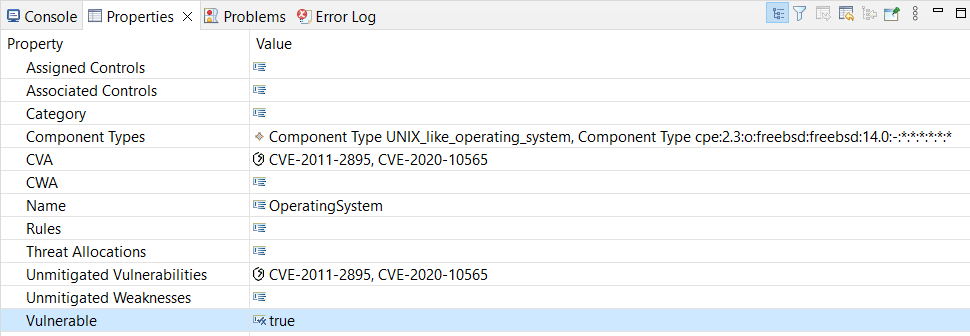}}
\caption{The OperatingSystem component's type mapping is set to include the new FreeBSD component type. The automated reasoning mechanism assesses the component as Vulnerable and lists unmitigated vulnerabilities.} \label{figOStype}
\end{figure}
\noindent

The vendor advisory for CVE-2020-10565\footnote{\url{https://svnweb.freebsd.org/ports?view=revision&revision=525916}, Accessed: 3/4/2025 } is to apply a specific patch. We capture this using \textit{rule2}. Similarly, the NIST NVD record for CVE-2011-2895 provides links to another patch, to solve this issue. We capture this using \textit{rule3}. Each of these newly added rules refers to newly identified security controls in the form of specific patches. Fig.~\ref{rule3} shows that our design space now includes two new rules and two new controls, and the details of \textit{rule3} as an example.

\begin{figure}[h]
\centering
\framebox{\includegraphics[scale=0.55]{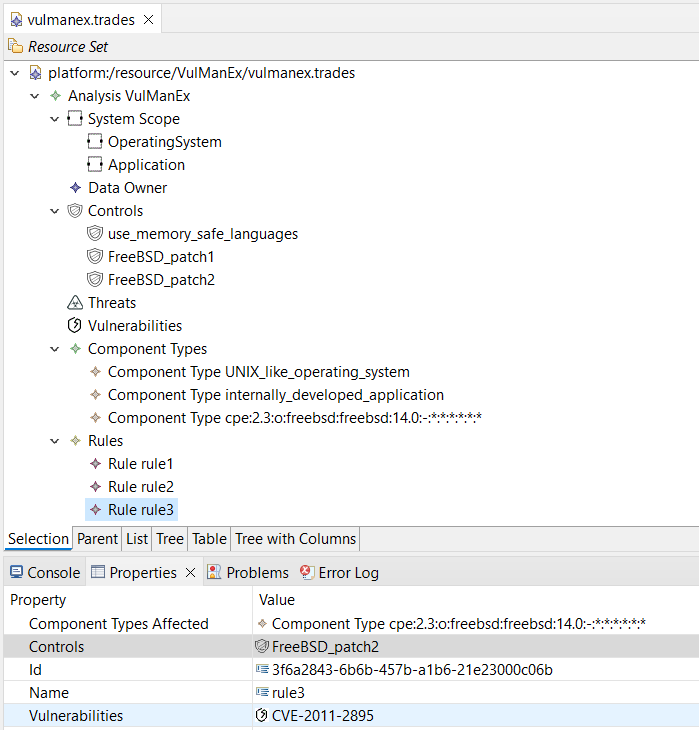}}
\caption{The system design model now includes two new controls representing available FreeBSD patches as well as two new rules to specify the use of each patch as a mitigation for a specific implementation vulnerability.} \label{rule3}
\end{figure}

We can capture a design decision to incorporate the two patches -- for example, while preparing a new deployment of our system -- into the system design model by associating the relevant security controls with the OperatingSystem component. Fig.~\ref{figOSpatched} shows the immediate results of the automated reasoning mechanism following such association: the OperatingSystem component is no longer vulnerable.
\begin{figure}[h]
\centering
\framebox{\includegraphics[scale=0.5]{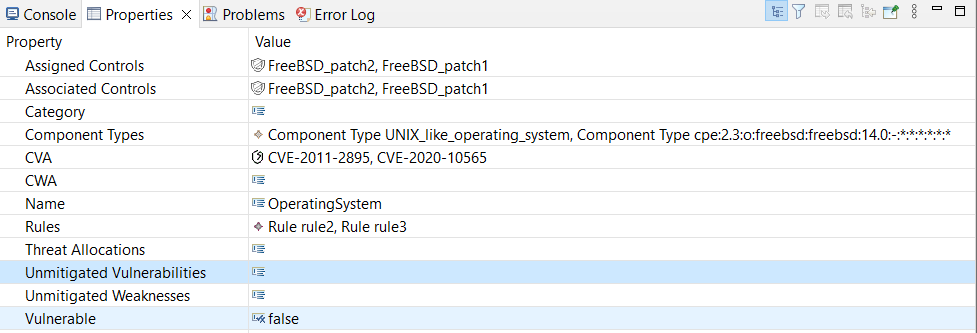}}
\caption{Once pertinent patches are assigned to the OperatingSystem component, the automated reasoning mechanism assesses that the component is no longer vulnerable.} \label{figOSpatched}
\end{figure}

As new implementation vulnerabilities emerge, new patches should be identified as security controls and new rules should address each specific vulnerability. Alternatively, we can address vulnerabilities by design, i.e., by eliminating the class of vulnerabilities, pending pertinent security controls. One such mitigation is to use a capability based addressing hardware (such as CHERI-Morello~\cite{cisa_memsafe}). We introduce this new security control into the knowledge base within our design environment, and specify a new rule -- \textit{rule4} -- to guide system designers in addressing CWE-119 in FreeBSD instances (Fig.~\ref{figCWERule}). We can then assign the new control to our OperatingSystem, instead of the previous patches. Fig.~\ref{figOSnotVul} shows this assignment as well as the assessment -- by the automated reasoning mechanism -- that the component is not vulnerable. This explicitly shows that the mechanism considers the mitigation of the CWE-119 mechanism vulnerability also as mitigation to the implementation vulnerabilities that manifest CWE-119: the potentially applicable vulnerabilities listed under the computed cVA mapping -- which is the implementation of the CVULNS(c) collection (as indicated in the previous section) -- do not appear under the Unmitigated Vulnerabilities, indicating that they are mitigated in the current system design.
\begin{figure}[h]
\centering
\framebox{\includegraphics[scale=0.45]{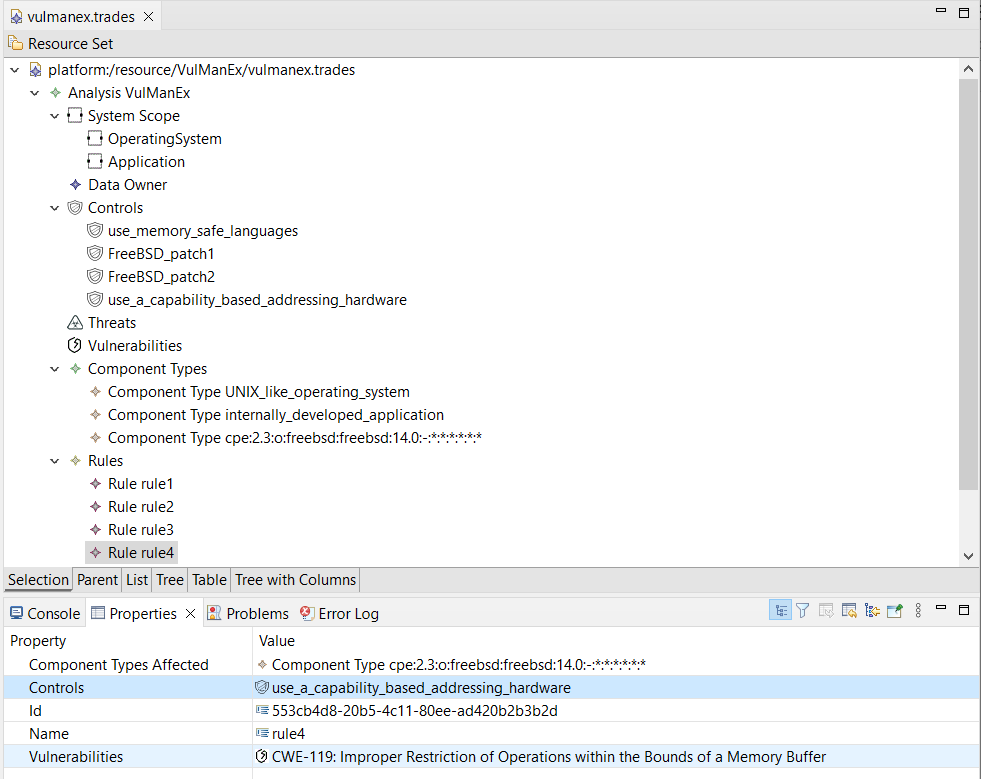}}
\caption{A new design rule prescribes the use of capability based addressing hardware as mitigation for the CWE-119 mechanism vulnerability in the context of FreeBSD.} \label{figCWERule}
\end{figure}
\FloatBarrier
\begin{figure}[h]
\centering
\framebox{\includegraphics[scale=1]{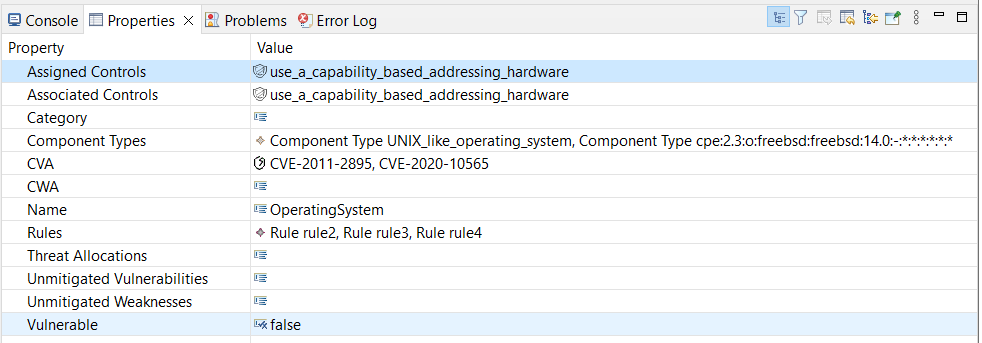}}
\caption{A new alternative design of the system includes an OperatingSystem that uses a capability based addressing hardware. Accordingly, the automated reasoning mechanism assesses that the FreeBSD OperatingSystem is not vulnerable to the collection of vulnerabilities (CVA).} \label{figOSnotVul}
\end{figure}
\FloatBarrier
\vspace{-10mm}
\section{Discussion}\label{sec:conclusion}Accounting for vulnerabilities in the design of systems requires careful
and rigorous consideration. In this paper, we have introduced formal
foundations to reason about the vulnerability posture of a system, and have
demonstrated a simple yet representative application. While our illustrative example is intentionally simple for clarity, it remains representative of design-related vulnerability management for two reasons: (1) our tool's knowledge base is populated with real-world vulnerability data, retrieved from the CWE and NVD databases; (2) the reasoning mechanism operates at the design level, based on design decisions (e.g., choosing a programming language for an internally developed application, selecting an operating system or an hardware platform, deploying a new version of a system with patches installed) as opposed to implementation details (e.g., specific code addressing a buffer beyond its boundaries). Accordingly, the example makes a valid, general case for adopting the automated reasoning mechanism for vulnerability management.

The suggested vulnerability management by design formalism is rooted in
well-established vulnerability management concepts, most notably the
concepts of \textit{component}, \textit{component type},
\textit{vulnerability} (in various levels of abstraction), and
\textit{control}. Consequently, the formalism is integrative. We have presented
an integration of the formalism into an open-source system security design
tool. While the integration is already fully functional, providing
automated reasoning capabilities, we are further integrating the reasoning
procedure and results into the diagrammatic representations and improving other user
experience aspects of the tool. Similarly, the formalism can be incorporated into other design and process management tools.

We are working towards adding additional design-related reasoning based on
formal properties and establishing their value in design contexts. For
example, another property can mandate the existence of sufficient
rules for addressing known vulnerabilities. A violation of such a property can trigger a security engineer to formulate additional rules, thereby
enriching the knowledge base that is available within the modelling
environment in support of more resilient designs. We are also considering separation between implementation vulnerabilities and mechanism vulnerabilities, to provide better quantitative
assessment of mitigation strategies and their coverage of existing and
future vulnerabilities.

While our formally grounded automated reasoning mechanism is scalable, we are well aware that manually specifying the models of the systems can be time consuming. For exercising the formalism and the automated reasoning capabilities at scale, further research can attempt to automate the generation of the models. A possible approach could be to use Software Bill of Materials records -- indicating the software components of products -- to populate the formal model. Once the formal model's definitions are in place, the automated reasoning can be applied without additional effort, as our TRADES Tool implementation demonstrates.

\begin{credits}
\subsubsection{\ackname} This work is funded by Innovate UK, grant number 75243, and by the ICO, Cybersecurity Institute of Occitanie, France.
\subsubsection{\discintname}
The authors have no competing interests to declare that are
relevant to the content of this article.
\end{credits}
%
%
%
\bibliographystyle{splncs04}
\bibliography{mybibliography}

\end{document}